\documentclass[12pt]{article}

\usepackage{amsmath}

\begin{document}

\bibliographystyle{unsrt}

\begin{titlepage}

\title{\bf An algebraic method for solving the SU(3) Gauss law}
\author{Antti Salmela\footnote{Email: Antti.Salmela@Helsinki.Fi} \\
\it Theoretical Physics Division \\ \it Department of Physical Sciences \\
\it P.O. Box 64, 00014 University of Helsinki, Finland}
\date{}

\maketitle

\thispagestyle{empty}

\begin{abstract}

A generalisation of existing SU(2) results is obtained. In particular, the source-free
Gauss law for SU(3)-valued gauge fields is solved using a non-Abelian analogue of the
Poincar\'e lemma. When sources are present, the colour-electric field is divided into
two parts in a way similar to the Hodge decomposition. Singularities due to coinciding
eigenvalues of the colour-magnetic field are also analysed.

\end{abstract}


\end{titlepage}

\pagebreak

\section{Introduction}

Gaining knowledge about the solutions of Gauss's law is important in view of the
central role that the law plays in quantising Yang--Mills theory. Usually the Gauss law
is ignored in the classical Hamiltonian formalism and then reintroduced at the quantum
level as a condition on the physical states. Yet in order to remove the redundant
degrees of freedom from the Hamiltonian we need a different approach. One way of
addressing this problem is to search for a method to parametrise the dynamical
variables of the theory so that Gauss's law is satisfied identically. The unconstrained
variables thus obtained will then describe the physical degrees of freedom of
Yang--Mills theory. In refs. \cite{gj} -- \cite{kp} a number of methods for working out
parametrisations of this kind are presented, but yet another approach was proposed some
years ago by Majumdar and Sharatchandra in ref. \cite{ms1}. They parametrised the
solutions of the SU(2) Gauss law
\begin{eqnarray}
&& \sum_{k=1}^3 \nabla_k(A) E_k = 0, \label{gsf} \\* && \nabla_k (A) = \partial_k + ig
[A_k (x),\, \cdot \, ] \nonumber
\end{eqnarray}
by expressing $E_k$ as a sum of a covariant curl and a gradient thus obtaining an SU(2)
generalisation of the Poincar\'e lemma. In order to make use of this decomposition in
QCD we need to generalise the results of ref. \cite{ms1} to SU(3), and it is the
purpose of this paper to provide such an extension. Hopefully, the parametrisation
could then serve as a starting point for developing Hamiltonian formalism according to
the lines sketched above. Besides Gauss's law, the decomposition might also be useful
in parametrising the non-Abelian generalisation of the Coulomb gauge
$$ \sum_{k=1}^3 \nabla_k(A) \dot{A}_k = 0 $$
proposed by Cronstr\"om \cite{cc}.

\section{SU(3) algebra}

I write every element of the SU(3) algebra in the form
$$A = \frac{1}{2} A^a \lambda_a,$$
where the $\lambda_a$'s stand for the Gell-Mann matrices
$$\lambda_a \lambda_b = \frac{2}{3} \delta_{ab} \, {\bf 1}_{3 \times 3} + \left( {d_{ab}}^c
+ i {f_{ab}}^c
\right) \lambda_c.$$
Summation over repeated indices is implied. An inner product
between two algebra elements is given by the Killing form
\begin{eqnarray*}
(A,B) &=& h_{ab} A^a B^b = 6 \, {\rm Tr} (AB), \\* h_{ab} &=& - {f_{ac}}^d {f_{bd}}^c =
\, 3 \, \delta_{ab}.
\end{eqnarray*}
I have chosen the convention where the inner product is positive definite. This inner
product defines a norm, which will be denoted by $|\cdot|$. The $d$ tensor can be used
to define a matrix-valued product
\begin{eqnarray*}
A*B &=& \frac{1}{2} {d_{ab}}^c A^a B^b \lambda_c \\*
 &=& \{A,B\} - \frac{1}{9} (A,B) \, {\bf 1}_{3 \times 3}.
\end{eqnarray*}
In addition to the Jacobi identity there exist several other identities involving the
structure constants of the algebra. They were worked out in ref. \cite{msw}:
\begin{subequations} \label{id}
\begin{eqnarray}
&&{f_{ea}}^d {d_{bc}}^e + {f_{eb}}^d {d_{ca}}^e + {f_{ec}}^d {d_{ab}}^e = 0
 \label{id1} \\
&&{f_{ea}}^b {f_{cd}}^e = \frac{2}{3} \left( \delta_{ac} {\delta^b}_d - \delta_{ad}
 {\delta^b}_c \right) + {d_{ac}}^e {d_{ed}}^b - {d_{ec}}^b {d_{ad}}^e \\
&&{d_{ad}}^e {d_{eb}}^c + {d_{bd}}^e {d_{ea}}^c + {d_{ed}}^c {d_{ab}}^e = \frac{1}{3}
\left( \delta_{ab} {\delta^c}_d + {\delta_b}^c \delta_{ad} + {\delta_a}^c \delta_{bd}
\right) \\
&& 3 \, {d_{ea}}^b {d_{cd}}^e = \delta_{ac} {\delta^b}_d + \delta_{ad} {\delta^b}_c -
{\delta_a}^b \delta_{cd} + {f_{ac}}^e {f_{de}}^b + {f_{ad}}^e {f_{ce}}^b.
\end{eqnarray}
\end{subequations}
These relations correspond to the matrix identities
\begin{subequations} \label{ego}
\begin{eqnarray}
&&[A*B,C] + [B*C,A] + [C*A,B] = 0 \label{ego1} \\
&&A*[B,C] + B*[A,C] + [C,A*B] = 0 \label{ego2} \\
&&{[A,[B,C]]} = \frac{2}{9} (A,B) C - \frac{2}{9} (A,C) B + C*(A*B) - B*(A*C) \label{ego3}
\\
&&A*(B*C) + B*(C*A) + C*(A*B) = \frac{1}{9} (A,B) C + \frac{1}{9} (B,C) A \nonumber \\*
&& \hspace{18.9em} + \frac{1}{9} (A,C) B \label{ego4} \\
&& 3 \, A*(B*C) = \frac{1}{3} (A,C) B + \frac{1}{3} (A,B) C - \frac{1}{3} (B,C) A +
[[A,C],B] \nonumber \\* && \hspace{7em} + [[A,B],C], \label{ego5}
\end{eqnarray}
\end{subequations}
equation (\ref{id1}) giving rise to both of the relations (\ref{ego1}) -- (\ref{ego2}).
Modifying the conventions of ref. \cite{msw} by some numerical factors I define two
invariants of the algebra
\begin{subequations} \label{inv}
\begin{eqnarray}
I_2(A) &=& |A|^2 \label{inv2} \\*
I_3(A) &=& (A,A*A) = 36 \det A. \label{inv3}
\end{eqnarray}
\end{subequations}
They remain unchanged under the adjoint action of the group
\begin{equation}
A \rightarrow \Omega^{\dagger} A \Omega, \qquad \Omega \in {\rm SU(3)}. \label{trans}
\end{equation}
Given a matrix $A$ one can define, following ref. \cite{msw}, another matrix
$\widehat{A}$
\begin{equation} \label{hat}
\widehat{A} = I_3(A) A - I_2(A) A*A
\end{equation}
with the properties
$$[A,\widehat{A}] = 0, \qquad (A,\widehat{A}) = 0.$$
This suggests that we should define a third invariant by
\begin{equation} \label{inv+}
I_8(A) = |\widehat{A}|^2 = I_2(A) \left( \frac{1}{9} I_2(A)^3 - I_3(A)^2 \right).
\end{equation}
Diagonalising $A$ with a transformation of the form (\ref{trans}) one can see that
$I_8$ vanishes if and only if $A$ has two coinciding eigenvalues. In the generic case
$I_8$ is strictly positive, though.

\section{Outline of solution}

In order to solve the Gauss law with sources
\begin{equation}\label{gauss}
\sum_{k=1}^3 \nabla_k(A) E_k = J_0
\end{equation}
I take an ansatz of the form
\begin{equation}\label{ans}
E_k = \sum_{l,m=1}^3 \varepsilon_{klm} \nabla_l(A) C_m + \nabla_k(A) \phi
\end{equation}
with the covariant derivative $\nabla_k(A)$ defined in equation (\ref{gsf}).
Analogously with the ordinary Hodge decomposition I define $\phi$ as a solution to the
covariant Poisson equation
\begin{equation} \label{pois}
\sum_{k=1}^3 \nabla_k^2(A) \phi = J_0.
\end{equation}
This equation has been analysed in detail in ref. \cite{as} and it proves to be
solvable for $\phi$ under certain fairly general conditions. Moreover, if $\phi$ is
assumed to vanish sufficiently rapidly at infinity, the solution is also unique.
Incidentally, Majumdar and Sharatchandra also included a covariant gradient term in
their ansatz for the source-free Gauss law \cite{ms1}, but their subsequent
calculations \cite{ms2} indicate that the gradient degrees of freedom are generically
redundant. In the Appendix I will discuss the question whether the ansatz (\ref{ans})
contains enough degrees of freedom to cover the space of colour-electric fields, but
for the moment I take the ansatz (\ref{ans}) for granted. Combining now Gauss's law
with the covariant divergence of equation (\ref{ans}) yields
\begin{equation}
\sum_{k=1}^3 ig [B_k,C_k] = 0, \label{ksum}
\end{equation}
where $B_k$ is the colour-magnetic field
$$ B_k = \sum_{l,m=1}^3 \varepsilon_{klm} \left( \partial_l A_m + \frac{1}{2} ig
[A_l,A_m] \right). $$ Equation (\ref{ksum}) could be solved by converting it into a
system of real-valued equations and applying standard tools of linear algebra such as
the Gauss elimination method. However, the  elimination procedure would give very
little insight into the algebraic nature of equation (\ref{ksum}) and the solution
obtained in this way would be complicated and formal. For this reason I prefer a less
straightforward method, which gives simpler solutions and makes algebraic features more
transparent. To begin with, let us parametrise the images of the commutators appearing
in equation (\ref{ksum}). More precisely, each commutator takes a matrix value
\begin{equation}\label{keq}
ig [B_k,C_k] = F_k,
\end{equation}
where $F_k$ must satisfy certain consistency conditions so that equation (\ref{keq})
can be solved for $C_k$. Making use of the following property of the inner product
$$ (X,i[B_k,C_k]) = -(i[B_k,X],C_k) $$
we see that $F_k$ must be orthogonal to all matrices that commute with $B_k$. I will
solve equation (\ref{keq}) properly in the next chapter, and it will turn out that in
the generic case the solvability conditions read
\begin{equation}\label{solv}
(F_k,B_k)=0, \qquad (F_k,\widehat{B}_k)=0,
\end{equation}
where $\widehat{B}_k$ is defined according to equation (\ref{hat}). The geometric
content of equations (\ref{ksum}) -- (\ref{solv}) becomes clearer if we start regarding
each matrix of the SU(3) algebra as an octet vector. The problem of parametrising the
solutions of equation (\ref{ksum}) is then reduced to parametrising all possible sets
of three vectors $F_k$ which satisfy the equation
\begin{equation}\label{fsum}
\sum_{k=1}^3 F_k = 0
\end{equation}
and the orthogonality conditions (\ref{solv}). This task is simplified by a suitable
choice of a basis for the SU(3) algebra. Generically, the following set of vectors will
serve as a basis:
\begin{equation}\label{basis}
\left\{ \begin{array}{ll}
 i[B_k,B_l], & k < l \\
 i[B_k,\widehat{B}_l] + i[\widehat{B}_k,B_l], & k < l \\
 \chi_1, \chi_2. & \end{array} \right.
\end{equation}
Here $\chi_1$ and $\chi_2$ are some vectors which are orthogonal to all of the six
vectors $B_k$ and $\widehat{B}_k$. We can define them as determinants
$$ \chi_j = \frac{1}{2} {\varepsilon_{a_1 \cdots a_6 b}}^c B_1^{a_1} \widehat{B}_1^{a_2}
B_2^{a_3} \widehat{B}_2^{a_4} B_3^{a_5} \widehat{B}_3^{a_6} \eta_j^b \lambda_c, \quad j=1,2 $$
where the $\eta_j$'s are some constant octet vectors. Taking $\eta_j$ parallel to some
Gell-Mann matrix $\lambda_a$ would reduce $\chi_j$ to a $7 \times 7$ determinant. To
see the linear independence of the set (\ref{basis}) let us consider the equation
\begin{equation}\label{indep}
i\sum_{k<l} a_{kl} [B_k,B_l] + i\sum_{k<l} \widehat{a}_{kl} ( [B_k,\widehat{B}_l] +
[\widehat{B}_k,B_l] ) + b_1 \chi_1 + b_2 \chi_2 = 0.
\end{equation}
Taking the inner product with respect to $B_m$ and $\widehat{B}_m$ leads to the
following pair of equations
$$ \left\{ \begin{array}{l}
 \! a_{kl} (B_m,i[B_k,B_l]) + \widehat{a}_{kl} (B_m,i[B_k,\widehat{B}_l] +
 i[\widehat{B}_k,B_l]) = 0 \\
 \! a_{kl} (\widehat{B}_m,i[B_k,B_l]) + \widehat{a}_{kl} (\widehat{B}_m,i[B_k,
 \widehat{B}_l] + i[\widehat{B}_k,B_l]) = 0
 \end{array} \right. $$
with $m \neq k \neq l.$ These equations have no nontrivial solutions if
\begin{eqnarray} \label{notr}
&& \Bigl( (B_m,i[B_k,B_l]) (\widehat{B}_m,i[B_k,\widehat{B}_l] + i[\widehat{B}_k,B_l])
\\*
&&\hspace{1em} - (\widehat{B}_m,i[B_k,B_l]) (B_m,i[B_k,\widehat{B}_l] +
i[\widehat{B}_k,B_l])  \Bigr) \neq 0. \nonumber
\end{eqnarray}
Generically this condition is satisfied, since none of the identities (\ref{ego})
implies that the expression inside the parentheses should vanish. It is also possible
to verify numerically, that is by assigning some test values to the vectors $B_k$, that
this expression does not vanish identically. In the same way we see that the remaining
coefficients $b_1$ and $b_2$ in equation (\ref{indep}) vanish if
\begin{equation} \label{eitr}
(\chi_1,\chi_1) (\chi_2,\chi_2) - \left[ (\chi_1,\chi_2) \right]^2 \neq 0.
\end{equation}
As before, this is generically satisfied because the l.h.s. does not vanish
identically. The linear independence of the set (\ref{basis}) thus proven in the
generic case, we use it as a basis for the vectors $F_k$,
\begin{equation} \label{fexp}
F_k = i\sum_{\substack{l=1 \\ l \neq k}}^3 \left( \alpha_{kl} [B_k,B_l] +
\widehat{\alpha}_{kl} ([B_k,\widehat{B}_l] + [\widehat{B}_k,B_l]) \right) + \beta_{1,k}
\chi_1 + \beta_{2,k} \chi_2.
\end{equation}
It should be noted that only six basis vectors are needed due to the orthogonality
conditions (\ref{solv}). Substituting now these expansions into equation (\ref{fsum})
gives the following relations:
\begin{eqnarray} \label{rela}
&&\alpha_{kl} - \alpha_{lk} = 0, \qquad \widehat{\alpha}_{kl} - \widehat{\alpha}_{lk} =
0, \nonumber \\*
&&\hspace{0.9em} \sum_{k=1}^3 \beta_{1,k} = 0, \qquad \sum_{k=1}^3
\beta_{2,k} = 0.
\end{eqnarray}
Let us finally count the number of degrees of freedom. Equation (\ref{rela}) states
that the matrices $\alpha$ and $\widehat{\alpha}$ are symmetric. Since four of the six
coefficients $\beta_{i,k}$ are independent, the total number of free variables is $2
\times 3 + 2 \times 2 = 10$. This is the number of coordinates needed to parametrise
three 6-dimensional vectors $F_k$ satisfying the 8-component equation (\ref{fsum}). We
have thus found all solutions to equations (\ref{solv}) -- (\ref{fsum}) in the generic
case, expressed in the form of expansion (\ref{fexp}) satisfying the relations
(\ref{rela}). Naturally there are nongeneric cases when either of the conditions
(\ref{notr}) -- (\ref{eitr}) fails and the set (\ref{basis}) becomes linearly
dependent. In those cases we must choose a different basis for the SU(3) algebra or at
least replace the ill-behaved vectors of the set (\ref{basis}) with linearly
independent ones. The method of solving equation (\ref{fsum}) remains the same even if
the basis is modified.

\section{Inverse of the commutator}

The vectors $F_k$ now known, it remains to solve equation (\ref{keq}) for the $C_k$'s.
Since the indices $k$ are fixed at this stage, I will omit them for a moment and
consider the equation
\begin{equation}\label{nok}
ig [B,C] = F.
\end{equation}
To obtain the solvability conditions for $F$ we must determine the zero modes of the
commutator. For that purpose, let us express the l.h.s. of equation (\ref{nok}) using
octet vector notation
$$ [B,C]^a = i {M^a}_c \, C^c, $$
where
$$ {M^a}_c = {f_{bc}}^a B^b. $$
The characteristic polynomial of $M$ becomes simpler to evaluate if we diagonalise $B$
by a suitable unitary transformation of the form (\ref{trans}),
\begin{equation}\label{diag}
\Omega^{\dagger} B \Omega = \frac{1}{2} b^3 \lambda_3 + \frac{1}{2} b^8 \lambda_8,
\qquad \Omega \in {\rm SU(3)}.
\end{equation}
On the other hand, this transformation can equivalently be implemented by an orthogonal
$8 \times 8$ matrix $O$
$$ \left( \Omega^{\dagger} B \Omega \right)^a = {O^a}_b B^b. $$
Since the structure constants ${f_{ab}}^c$ transform as a tensor, we have
\begin{eqnarray*}
{M^a}_c &=& {O_b}^a {O^d}_c {O^e}_f {f_{ed}}^b \left( {O_3}^f \, b^3 + {O_8}^f \, b^8
 \right) \\*
&=& {O_b}^a {\widetilde{M}^b}_{\,\,\,\, d} {O^d}_c, \\
{\widetilde{M}^b}_{\,\,\,\, d} &=& b^3 {f_{3d}}^b + b^8 {f_{8d}}^b.
\end{eqnarray*}
A straightforward calculation now gives
\begin{eqnarray*}
&& \det \left( M - x \, {\bf 1}_{8 \times 8} \right) =  \det \left( \widetilde{M} - x
\, {\bf 1}_{8 \times 8} \right) \\*
&& \hspace{1.4em} = \, x^2 \Bigl\{ x^6 + \frac{3}{2} \left[ (b^3)^2 + (b^8)^2 \right]
x^4 + \frac{9}{16} \left[ (b^3)^2 + (b^8)^2 \right]^2 x^2 \\
&& \hspace{4.1em} + \frac{1}{16} (b^3)^2 \left[ (b^3)^2 - 3 (b^8)^2 \right]^2 \Bigr\}.
\end{eqnarray*}
With the help of the invariants (\ref{inv2}) and (\ref{inv+}),
\begin{eqnarray*}
I_2(B) &=& 3 \left[ (b^3)^2 + (b^8)^2 \right] \\* I_8(B) &=& 9 (b^3)^2 \left[ (b^3)^2 +
(b^8)^2 \right] \left[ (b^3)^2 - 3 (b^8)^2 \right]^2,
\end{eqnarray*}
the characteristic polynomial can be written as
$$ \det \left( M - x \, {\bf 1}_{8 \times 8} \right) = x^2 \left[ x^2 \left( x^2 +
\frac{1}{4} I_2 \right)^2  +
 \frac{1}{48} \frac{I_8}{I_2} \right]. $$
The commutators thus fall into three classes according to the number of zero modes:
\begin{enumerate}
\item $I_2 > 0$, $I_8 > 0$ \hfill \break This is the generic case, when all eigenvalues
of $B$ are distinct. The zero modes are given by $B$ itself and the matrix
$\widehat{B}$ defined in equation (\ref{hat}). \item $I_2 > 0$, $I_8 = 0$ \hfill \break
In this case $B$ is nonvanishing but has two coinciding eigenvalues. One can conjugate
$B$ into the direction of $\lambda_8$:
\begin{equation}\label{bform}
B = \Omega \left( \frac{1}{2} b^8 \lambda_8 \right) \Omega^{\dagger}.
\end{equation}
There are four zero modes, which are obtained by conjugating all the Gell-Mann matrices
that commute with $\lambda_8$, i.e. they have the form
$$ \Omega \left( \frac{1}{2} \lambda_a \right)  \Omega^{\dagger}, \qquad a=1,2,3,8.$$
\item $I_2 = 0$, $I_8 = 0$ \hfill \break
This case is trivial, because $B$ vanishes.
\end{enumerate}
Let us now solve equation (\ref{nok}) in the generic case. There are solutions only if
$F$ is orthogonal to the zero modes of the commutator, i.e.
$$ (F,B)=0, \qquad (F,\widehat{B})=0. $$
Introducing a projection operator
\begin{equation}\label{projo}
\Pi(F) = F - \frac{1}{I_2(B)} (B,F) B - \frac{1}{I_8(B)} (\widehat{B},F) \widehat{B},
\end{equation}
the conditions on $F$ can also be expressed as a single equation
$$ F = \Pi(F). $$
Equation (\ref{nok}) can thus be replaced by
\begin{equation}\label{preq}
ig [B,C] = \Pi(F).
\end{equation}
The general form of the solution $C$ is
\begin{equation}\label{form}
C^a = {t^a}_b F^b,
\end{equation}
where the tensor $t$ depends only on $B$, because $C$ must be linear in $F$. The basis
for all such tensors was given in ref. \cite{msw}, and substituting it into equation
(\ref{form}) yields the following ansatz
\begin{eqnarray*}
C &=& a_1 F + a_2 B*F + a_3 \widehat{B}*F + a_4 [B,F] + a_5 [\widehat{B},F] + a_6
B*[\widehat{B},F] \nonumber \\*
&& + a_7 B + a_8 \widehat{B}.
\end{eqnarray*}
Using identities (\ref{ego}) the commutator of this expression becomes
\begin{eqnarray}\label{kommu}
[B,C] &=& \left( \frac{I_2}{6} a_4 - \frac{I_8}{12 \, I_2} a_6 \right) F + \left(
\frac{3 \, I_3}{2 \, I_2} a_4 -
\frac{3 \, I_8}{2 \, I_2^2} a_5 \right) B*F \nonumber \\
&& - \left( \frac{3}{2 \, I_2} a_4 + \frac{3 \, I_3}{2 \, I_2} a_5 + \frac{I_2}{12} a_6
\right) \widehat{B}*F \nonumber \\
&& + \left( a_1 + \frac{I_3}{2 \, I_2} a_2 - \frac{I_8}{I_2^2} a_3 \right) [B,F] \\
&& -  \left( \frac{1}{2 \, I_2} a_2 + \frac{I_3}{I_2} a_3 \right) [\widehat{B},F] - a_3
\, B*[\widehat{B},F] \nonumber \\
&& - \frac{1}{3} a_4 \, (B,F) B + \frac{1}{6 \, I_2} a_6 \, (\widehat{B},F) \widehat{B}
\nonumber \\
&& - \left( \frac{1}{6} a_5 + \frac{I_3}{12 \, I_2} a_6 \right) \left( (\widehat{B},F)
B + (B,F) \widehat{B} \right). \nonumber
\end{eqnarray}
Inserting expansions (\ref{kommu}) and (\ref{projo}) into equation (\ref{preq}) and
equating terms of the same form determines six of the coefficients $a_i$,
\begin{eqnarray*}
&& a_1 = 0, \qquad a_2 = 0, \qquad a_3 = 0, \\*
&& a_4 = - \frac{i}{g} \frac{3}{I_2},
\qquad a_5 = - \frac{i}{g} \frac{3 \, I_3}{ I_8}, \qquad a_6 = \frac{i}{g} \frac{6 \,
I_2}{I_8}.
\end{eqnarray*}
Hence, the solution to equation (\ref{preq}) is
\begin{equation}\label{solu}
C = - \frac{3i}{g} \left( \frac{1}{I_2} [B,F] + \frac{I_3}{I_8} [\widehat{B},F] -
\frac{2 \, I_2}{I_8} B*[\widehat{B},F] \right) + a_7 B + a_8 \widehat{B}.
\end{equation}
This formula becomes singular when $I_8$ tends to zero. In this limit $B$ can be
written in the form (\ref{bform}). The orthogonality conditions on $F$ are
\begin{equation} \label{excon}
\left( \Omega \left( \frac{1}{2} \lambda_a \right)  \Omega^{\dagger},F \right) = 0,
\qquad a = 1,2,3,8.
\end{equation}
When these requirements are fullfilled, it is straightforward to see that the following
expression satisfies equation (\ref{nok}):
\begin{equation}\label{except}
C = - \frac{4i}{g} \frac{1}{I_2} [B,F]  + \Omega \Bigl( \frac{1}{2} a^1 \lambda_1 +
\frac{1}{2} a^2 \lambda_2  + \frac{1}{2} a^3 \lambda_3 + \frac{1}{2} a^8 \lambda_8
\Bigr) \Omega^{\dagger}.
\end{equation}
Formulas (\ref{solu}) and (\ref{except}) thus solve the commutator equation (\ref{nok})
in the two nontrivial cases.

\section{Results}

We are now ready to write down the general solution to equation (\ref{ksum}). In the
generic case the expansions (\ref{fexp}) parametrise all possible values for the
commutators (\ref{keq}). Substituting these expansions into equation (\ref{solu}) and
simplifying the result with the identities (\ref{ego}) yields the solution
\begin{eqnarray}\label{res}
C_k &=& \frac{1}{g} \sum_{\substack{l=1 \\ l \neq k}}^3 \Bigl[ (\alpha_{kl} + I_3^{(k)}
\widehat{\alpha}_{kl}) B_l + \widehat{\alpha}_{kl} \left( \widehat{B}_l - 2 I_2^{(k)}
B_k * B_l \right) \Bigr] \nonumber \\* && + \frac{1}{g} \sum_{j=1}^2 \beta_{j,k}
\psi_{j,k} + \gamma_k B_k + \widehat{\gamma}_k \widehat{B}_k,
\end{eqnarray}
where
\begin{eqnarray} \label{psi}
\Pi_k (\psi_{j,k}) &=& -3i \Bigl( \frac{1}{I_2^{(k)}} [B_k, \chi_j] +
\frac{I_3^{(k)}}{I_8^{(k)}} [\widehat{B}_k, \chi_j] - \frac{2 \, I_2^{(k)}}{I_8^{(k)}}
B_k *[\widehat{B}_k,\chi_j] \Bigr), \nonumber \\* I_i^{(k)} &\equiv& I_i(B_k)
\end{eqnarray}
and $\Pi_k$ stands for the projection operator of equation (\ref{projo}). The symmetry
relations (\ref{rela}) must hold, while the zero mode coefficients $\gamma_k$ and
$\widehat{\gamma}_k$ are arbitrary. I have left unsimplified those parts of the
solution which correspond to the two vectors $\chi_j$. Of course, simplifications can
be performed using the results of ref. \cite{ammp}. In particular, it is shown there
how the eighth rank permutation symbol can be expressed in a form involving only the
structure constants ${f_{ab}}^c$ and ${d_{ab}}^c$. Constructing all possible sixth rank
tensors that are antisymmetric in five indices and contracting them with the vectors
$B_l$, $\widehat{B}_l$ ($l \neq k$) and $\eta_j$ would give us the vectors needed to
reduce the expression (\ref{psi}). Unfortunately there is such a large number of these
tensors that the resulting formula would be unduly long. The unsimplified formula
(\ref{psi}) is the shortest expression I have been able to obtain. Still there is a
relatively simple formula at hand if we diagonalise $B_k$ with a transformation of the
form (\ref{diag}). Making use of the fact that the eighth rank permutation symbol
transforms as a tensor under the adjoint action (\ref{trans}) leads to
\begin{equation} \label{sipsi}
\Pi_k (\psi_{j,k}) = \frac{1}{\sqrt{3}} I_2^{(k)} c^{ab}_k \Delta_{38b}^{(j,k)} \,
\Omega_k \! \left( \frac{1}{2} \lambda_a \right) \! \Omega^{\dagger}_k,
\end{equation}
where
\begin{eqnarray*}
\Delta_{abc}^{(j,k)} &=& \varepsilon_{ab d_1 \cdots d_5 c} \left( \Omega^{\dagger}_k B_l
\Omega_k \right)^{d_1} \left( \Omega^{\dagger}_k \widehat{B}_l \Omega_k \right)^{d_2} \left(
\Omega^{\dagger}_k B_m \Omega_k \right)^{d_3} \\*
&& \hspace{2.7em} \times \left( \Omega^{\dagger}_k \widehat{B}_m \Omega_k \right)^{d_4}
\left( \Omega^{\dagger}_k \eta_j \Omega_k \right)^{d_5}, \qquad l \neq m \neq k, \\
c^{12}_k &=& - c^{21}_k = \frac{1}{3} [(b_k^3)^2 - 3(b_k^8)^2] \\
c^{45}_k &=& - c^{54}_k = \frac{2}{3} b_k^3 (b_k^3 - \sqrt{3} \, b_k^8) \\
c^{67}_k &=& - c^{76}_k = -\frac{2}{3} b_k^3 (b_k^3 + \sqrt{3} \, b_k^8)
\end{eqnarray*}
and all the other components of the matrix $c_k$ vanish.

In order to avoid singularities in the limit when two eigenvalues of $B_k$ coincide we
must find a way to regularise the vectors $\psi_{j,k}$. This singularity is present in
equation (\ref{solu}), but it does not mean that the solution (\ref{res}) would have to
be singular. Actually, even the first six vectors of the basis (\ref{basis}), when
inserted into equation (\ref{solu}), produced singular terms, but these terms were
proportional to the zero modes $B_k$ and $\widehat{B}_k$. The singularities could then
be removed by redefining the zero mode coefficients $\gamma_k$ and
$\widehat{\gamma}_k$, and the same procedure can also be applied to the vectors
$\psi_{j,k}$. Specifically, let us define
\begin{equation} \label{repsi}
\psi_{j,k} = - \frac{1}{2 \sqrt{3}} \, I_2^{(k)} \tilde{c}^a_{j,k} \, \Omega_k \! \left(
\frac{1}{2} \lambda_a \right) \! \Omega^{\dagger}_k,
\end{equation}
with
\begin{eqnarray*}
\tilde{c}^a_{j,k} &=& \frac{1}{3} [(b_k^3)^2 - 3(b_k^8)^2] \, \varepsilon_{123}^{abc}
\Delta_{b8c}^{(j,k)} \\
 && -\frac{2}{3} b_k^3 (b_k^3 - \sqrt{3} \, b_k^8) \left( \frac{1}{2}
 \varepsilon_{453}^{abc} + \frac{\sqrt{3}}{2} \varepsilon_{458}^{abc} \right) \left(
 \frac{\sqrt{3}}{2} \Delta_{b3c}^{(j,k)} - \frac{1}{2} \Delta_{b8c}^{(j,k)} \right) \\
 && -\frac{2}{3} b_k^3 (b_k^3 + \sqrt{3} \, b_k^8) \left( -\frac{1}{2}
 \varepsilon_{673}^{abc} + \frac{\sqrt{3}}{2} \varepsilon_{678}^{abc} \right) \left(
 -\frac{\sqrt{3}}{2} \Delta_{b3c}^{(j,k)} - \frac{1}{2} \Delta_{b8c}^{(j,k)} \right),
\end{eqnarray*}
where $\varepsilon_{ijk}^{abc}$ stands for the three-dimensional permutation symbol
with indices taking the values $i$, $j$ and $k$. Equations (\ref{repsi}) and
(\ref{sipsi}) are equivalent apart from terms which are proportional to the zero modes.
We can now pass to the limit when two eigenvalues of $B_k$ coincide. Let us assume that
the eigenvalues are ordered so that
$$ b_k^3 + \frac{1}{\sqrt{3}} \, b_k^8 \geq -b_k^3 + \frac{1}{\sqrt{3}} \, b_k^8 \geq
-\frac{2}{\sqrt{3}} \, b_k^8, $$
which means that the largest eigenvalues coincide in the limit $b_k^3 \rightarrow 0$.
In this limit the vectors $\psi_{j,k}$ are reduced to
\begin{equation} \label{su2}
\psi_{j,k} \rightarrow \frac{1}{6 \sqrt{3}} \left( I_2^{(k)} \right)^2
\varepsilon_{123}^{abc} \Delta_{b8c}^{(j,k)} \, \Omega_k \! \left( \frac{1}{2} \lambda_a
\right) \! \Omega^{\dagger}_k.
\end{equation}
In order to show that this expression is single-valued we must verify that it is
invariant under transformations which also leave $B_k$ invariant. As $B_k$ now takes
the form (\ref{bform}), the transformations in question are the SU(2)$\times$U(1)
reparametrisations of the matrix $\Omega_k$ defined by
\begin{eqnarray} \label{repa}
&& \Omega_k \rightarrow \Omega_k \, \omega, \\
&& \omega = \exp \left[ \frac{i}{2} ( \theta^1 \lambda_1 + \theta^2 \lambda_2 +
\theta^3 \lambda_3 + \theta^8 \lambda_8 ) \right]. \nonumber
\end{eqnarray}
These transformations take
\begin{eqnarray*}
&&\lambda_a \rightarrow {P_a}^{a'} \, \lambda_{a'} \\
&&\Delta_{b8c}^{(j,k)} \rightarrow {P_b}^{b'} {P_c}^{c'} \Delta_{b'8c'}^{(j,k)}, \\
&&{P_a}^{a'} = \left[ \omega \left( \frac{1}{2} \lambda_a \right) \omega^{\dagger}
\right]^{a'},
\end{eqnarray*}
and as a result
\begin{eqnarray*}
&& \psi_{j,k} \rightarrow \left( \det_{3 \times 3} P \right) \, \psi_{j,k}, \\
&& \det_{3 \times 3} P = \varepsilon_{123}^{abc} {P_a}^{1} {P_b}^{2} {P_c}^{3}.
\end{eqnarray*}
Since $\det_{3 \times 3} P = 1$ for transformations of the form (\ref{repa}), the
solution (\ref{su2}) is invariant. Although this result was derived in the case when
the largest eigenvalues of $B_k$ coincide, the form of the solution (\ref{repsi}) makes
it evident that $\psi_{j,k}$ is really single-valued regardless of which SU(2) subgroup
survives in the limit of coinciding eigenvalues. So there will be no singularities in
the fields $C_k$ in equation (\ref{res}).

So far we have found out that equation (\ref{ksum}) possesses solutions which are
regular everywhere in space. Yet it is possible that there might be some physically
relevant degrees of freedom residing at points where two eigenvalues of $B_k$ coincide
and that we should search for singular solutions to equation (\ref{ksum}) in order to
detect these degrees of freedom. In fact, it is widely believed that there are
singularities with local monopole-like behaviour in the vicinity of points where two
eigenvalues coincide. Usually these singularities arise as a result of gauge fixing
\cite{th}, but here they could emerge in connection with special solutions to equation
(\ref{ksum}). To determine such solutions we need to modify the basis (\ref{basis})
slightly. I will consider the case when one component of the colour-magnetic field, say
$B_3$, has coinciding eigenvalues. $B_3$ takes the form (\ref{bform}) and in particular
$\widehat{B}_3 = 0$. (I am not assuming that the eigenvalues should be ordered this
time.) Yet the first six vectors of the set (\ref{basis}) remain generically indepent
while $\chi_1$ and $\chi_2$ vanish. Since $F_3$ in equation (\ref{keq}) now has to
satisfy four orthogonality conditions according to equation (\ref{excon}), we see that
the vectors $\chi_j$ should be replaced by two vectors $\widetilde{\chi}_j$ which are
orthogonal to the space spanned by the set $\{ B_1, \widehat{B}_1, B_2, \widehat{B}_2
\}$. We can take
$$ \left\{ \begin{array}{l}
\widetilde{\chi}_1 = i[\widehat{B}_1,\widehat{B}_2] \\
\widetilde{\chi}_2 = [B_1,B_2]*[\widehat{B}_1,\widehat{B}_2] -
[B_1,\widehat{B}_2]*[\widehat{B}_1,B_2].
\end{array} \right. $$
The expansion for $F_k$ now takes the form of equation (\ref{fexp}) satisfying the
relations (\ref{rela}) with the obvious substitutions
$$ \left\{ \begin{array}{ll} \chi_j \rightarrow \widetilde{\chi}_j &\\
\beta_{j,k} \rightarrow \widetilde{\beta}_{j,k}, & \quad k=1,2 \\
\beta_{j,3} \rightarrow 0. & \end{array} \right. $$
Inserting these expansions into
equations (\ref{solu}) and (\ref{except}) leads to the solution
\begin{eqnarray}\label{excres}
C_k &=& \frac{1}{g} \sum_{\substack{l=1 \\ l \neq k}}^3 \Bigl[ (\alpha_{kl} + I_3^{(k)}
\widehat{\alpha}_{kl}) B_l + \widehat{\alpha}_{kl} \left( \widehat{B}_l - 2 I_2^{(k)}
B_k * B_l \right) \Bigr] \nonumber \\* && + \frac{1}{g} \sum_{j=1}^2
\widetilde{\beta}_{j,k} \widetilde{\psi}_{j,k} +
\gamma_k B_k + \widehat{\gamma}_k \widehat{B}_k, \qquad k = 1,2 \nonumber \\
C_3 &=& \frac{1}{g} \sum_{l=1}^2  \left( \alpha_{3l} B_l + \widehat{\alpha}_{3l}
\widehat{B}_l \right) \\* && + \Omega_3 \Bigl( \frac{1}{2} \widetilde{\gamma}_3^1
\lambda_1 + \frac{1}{2} \widetilde{\gamma}_3^2 \lambda_2 + \frac{1}{2}
\widetilde{\gamma}_3^3 \lambda_3 + \frac{1}{2} \widetilde{\gamma}_3^8 \lambda_8 \Bigr)
\Omega^{\dagger}_3. \nonumber
\end{eqnarray}
Here the coefficients $\widetilde{\gamma}_3^a$ are arbitrary and $\Omega_3$ is a matrix
which diagonalises $B_3$. The vectors $\widetilde{\psi}_{j,k}$ are defined as in
equation (\ref{psi}), replacing only $\chi_j \rightarrow \widetilde{\chi}_j$. Let us
now compare the two solutions (\ref{res}) and (\ref{excres}) at points where two
eigenvalues of $B_3$ coincide. As 't Hooft mentioned in ref. \cite{th}, this takes
place at isolated points in three-dimensional space for generic magnetic fields. At
such points the vectors $\psi_{j,1}$ and $\psi_{j,2}$ vanish, which corresponds to
setting $\widetilde{\beta}_{j,k} = 0$ in equation (\ref{excres}). Equating the
$C_3$-components of formulas (\ref{res}) and (\ref{excres}) with the help of equation
(\ref{su2}) determines the coefficients $\widetilde{\gamma}_3^a$ as functions of
$\beta_{j,3}$, $\gamma_3$ and $\widehat{\alpha}_{3l}$. However, since
$\widehat{\alpha}_{3l}$ already appears in the first term of equation (\ref{excres}),
there are effectively only three arbitrary parameters determining four unknown
coefficients and accordingly, there is one more degree of freedom in the solution
(\ref{excres}) which is not present in the formula (\ref{res}). In all, there are thus
three degrees of freedom in the exceptional solution (\ref{excres}) which cannot be
obtained by taking the limit of equation (\ref{res}). This leaves the door open for
accepting singular solutions to equation (\ref{ksum}). In that case, though, equation
(\ref{pois}) should be replaced by
$$ \sum_{k=1}^3 \nabla_k^2(A) \phi = J_0 \, - \!\! \sum_{k,l,m=1}^3 \! \varepsilon_{klm}
\partial_k \partial_l C_m $$
to compensate for the possibility that the second weak derivatives of $C_m$ do not
commute.

\section{Conclusions}

I have presented here a method by which the Gauss law (\ref{gauss}) can be solved in
the case of the SU(3) algebra using the ansatz (\ref{ans}). The fact that the l.h.s. of
the consistency equation (\ref{ksum}) depends on the commutator properties of the
colour-magnetic field divides the solutions into different classes. I have written down
the source-free part of the solution explicitly in the generic case when the set
(\ref{basis}) is linearly independent and in the case when one component of the
colour-magnetic field has coinciding eigenvalues. Although the SU(2) solution of ref.
\cite{ms1} was simple, its SU(3) generalisation (\ref{res}) is much more complicated.
The vectors $\chi_j$ of the basis (\ref{basis}) are mostly responsible for the
complexity, and unfortunately I see no way out of this problem. I could replace the
$\chi_j$'s by vectors which would be easier to invert with the formula (\ref{solu}),
e.g. by
$$ i[\widehat{B}_1,\widehat{B}_2], i[\widehat{B}_1,\widehat{B}_3], $$
but then the orthogonality conditions (\ref{solv}) for $F_2$ and $F_3$ would lead to
complicated relations between the expansion coefficients. So there seems to be some
kind of "conservation of trouble" inherent in this problem. Anyway, it is interesting
that the fields $C_k$ may have singularities at points where one component of the
colour-magnetic field possesses two coinciding eigenvalues. No explicit gauge fixing is
needed to detect this singularity as it becomes apparent whenever one tries to solve
equation (\ref{ksum}). The method of solving this equation could also be generalised to
higher dimensional SU(N) algebras in a straightforward manner, but the results would
undoubtedly be even more complicated.

\appendix
\setcounter{equation}{0}
\renewcommand{\theequation}{A.\arabic{equation}}

\section*{Appendix: Motivation for the generalised \\ Hodge decomposition}

In order to prove that the space of colour-electric fields can be parametrised with the
ansatz (\ref{ans}) it is sufficient to show that the equation
\begin{eqnarray} \label{lapla}
E_k &=& \sum_{l=1}^3 \Bigl( \nabla_l^2(A) \Phi_k - ig [G_{kl},\Phi_l] \Bigr), \\
G_{kl} &=& \partial_l A_k - \partial_k A_l - ig [A_k,A_l] \nonumber \\*
 &=& - \sum_{m=1}^3 \varepsilon_{klm} B_m \nonumber
\end{eqnarray}
can be solved for the field $\Phi_k$. Making use of the identity
\begin{eqnarray*}
\sum_{l=1}^3 \Bigl( \nabla_l^2(A) \Phi_k - ig [G_{kl},\Phi_l] \Bigr) &=& -
\sum_{l,m=1}^3 \varepsilon_{klm} \nabla_l(A) \sum_{p,q=1}^3 \varepsilon_{mpq}
\nabla_p(A) \Phi_q \\*
&& + \nabla_k(A) \sum_{l=1}^3 \nabla_l(A) \Phi_l
\end{eqnarray*}
we see that equation (\ref{lapla}) then takes the form of equation (\ref{ans}) with
\begin{subequations} \label{konst}
\begin{eqnarray}
C_m &=& - \sum_{p,q=1}^3 \varepsilon_{mpq} \nabla_p(A) \Phi_q, \label{konst1} \\*
\phi
&=& \sum_{l=1}^3 \nabla_l(A) \Phi_l. \label{konst2}
\end{eqnarray}
\end{subequations}
I am not trying to solve equation (\ref{lapla}) here, but it seems fairly obvious that
a solution exists. In a finite volume this equation can be converted into an integral
equation after choosing suitable boundary conditions so that the ordinary Laplacian
$$\Delta = \sum_{l=1}^3 \partial_l^2$$
has a unique inverse. The resulting integral equation can then be set into the form of
a Fredholm equation and solved, at least formally, using the well-known Fredholm
formulas. In  infinite space this procedure would require that the fields $E_k$ and the
gauge potentials $A_k$ decay sufficiently rapidly at infinity.

Equation (\ref{lapla}) shows that the ansatz (\ref{ans}) contains one redundant SU(3)
algebra -valued field, because in general there are no relations like equations
(\ref{konst}) among $C_k$ and $\phi$. This gives rise to a heuristic argument in favour
of the choice (\ref{pois}) for the field $\phi$, since we seem to be free to fix one
field component at will. In order to be more exact we should investigate whether the
space of colour-electric fields with vanishing covariant divergences can be
parametrised with the covariant curl ansatz
\begin{equation}\label{frans}
\widetilde{E}_k = \sum_{l,m=1}^3 \varepsilon_{klm} \nabla_l(A) C_m,
\end{equation}
where
\begin{eqnarray*}
&& \sum_{k=1}^3 \nabla_k(A) \widetilde{E}_k = 0, \\
&& \widetilde{E}_k = E_k - \nabla_k(A) \phi.
\end{eqnarray*}
In ref. \cite{ms2} Majumdar and Sharatchandra considered an SU(2) equation of the form
(\ref{frans}) with $\widetilde{E}_k = 0$ and presented a method for obtaining a formal
solution. Using the consistency condition (\ref{ksum}) they eliminated $C_3$ and
converted the remaining equations into the form of a Cauchy problem with initial data
given on the plane $x_3=0$. They showed that a formal solution to the Cauchy problem
can be constructed in a certain generic case as a power series near the initial plane
$x_3=0$. Unfortunately there is an error in their reasoning concerning the convergence
of the power series. Namely, they try to apply the Cauchy--Kovalevskaya theorem to
equations of the form
$$ \partial_3 C_j = G[C_1, C_2, \{A_k\}], \qquad j=1,2 $$
where the functional $G$ depends on second order derivatives of $C_1$ and $C_2$ with
respect to $x_1$ and $x_2$. In this case the Cauchy--Kovalevskaya theorem is not valid
and the formal solution does not necessarily converge. The method of ref. \cite{ms2}
would be easy to generalise to the case of equation (\ref{frans}), but the formal
solution might be mathematically meaningless. Anyway, this method hints that equation
(\ref{frans}) can generically be solved for the $C_k$'s and accordingly, that the space
of sufficiently regular colour-electric fields with vanishing covariant divergences can
be parametrised with the ansatz (\ref{frans}). From the mathematical point of view this
problem is still open.

\section*{Acknowledgements}

I would like to thank professor C. Cronstr\"om for his guidance during the preparation
of this paper. This research was supported by the Academy of Finland.


\begin{thebibliography}{99}

\bibitem{gj} J. Goldstone and R. Jackiw, Phys. Lett. B {\bf 74}, 81--84 (1978)

\bibitem{bg} V. Baluni and B. Grossman, Phys. Lett. B {\bf 78}, 226--230 (1978)

\bibitem{iksf} A. G. Izergin, V. E. Korepin, M. A. Semenov-Tian-Shansky and L. D.
Faddeev, Theor. Math. Phys. {\bf 38}, 1--9 (1979)

\bibitem{cmt} M. Creutz, I. J. Muzinich and T. N. Tudron, Phys. Rev. D {\bf 19},
531--539 (1979)

\bibitem{h} M. B. Halpern, Phys. Rev. D {\bf 19}, 517--530 (1979)

\bibitem{bfh} M. Bauer, D. Z. Freedman and P. E. Haagensen, Nucl. Phys. B {\bf 428},
147--168 (1994), hep-th/9405028

\bibitem{hj} P. E. Haagensen and K. Johnson, Nucl. Phys. B {\bf 439}, 597--616 (1995),
hep-th/9408164

\bibitem{s} R. Schiappa, Nucl. Phys. B {\bf 517}, 462--484 (1998), hep-th/9704206

\bibitem{n} H. Nachbagauer, Phys. Rev. D {\bf 52}, 3672--3678 (1995), hep-ph/9503244

\bibitem{cbh} L. Chen, M. Belloni and K. Haller, Phys. Rev. D {\bf 55}, 2347--2366
(1997), hep-ph/9609507

\bibitem{r} H. Reinhardt, Nucl. Phys. B {\bf 503}, 505--529 (1997), hep-th/9702049

\bibitem{kp} A. M. Khvedelidze and H.-P. Pavel, Phys. Rev. D {\bf 59}, 105017 (1999),
hep-th/9808102

\bibitem{ms1} P. Majumdar and H. S. Sharatchandra, Phys. Rev. D {\bf 58},
067702 (1998), hep-th/9804128

\bibitem{cc} C. Cronstr\"om, Acta Phys. Slov. {\bf 50}, 369--379 (2000), hep-th/9906184

\bibitem{msw} A. J. Macfarlane, A. Sudbery and P. H. Weisz, Commun. Math. Phys.
{\bf 11}, 77--90 (1968)

\bibitem{as} A. Salmela, math-ph/0211024, HIP-2002-59/TH, Helsinki Institute of Physics,
2002

\bibitem{ms2} P. Majumdar and H. S. Sharatchandra, Phys. Rev. D {\bf 63},
067701 (2001), hep-th/9804091

\bibitem{ammp} J. A. de Azc\'arraga, A. J. Macfarlane, A. J. Mountain and J. C.
P\'erez Bueno, Nucl. Phys. B {\bf 510}, 657--687 (1998), physics/9706006

\bibitem{th} G. 't Hooft, Nucl. Phys. B {\bf 190}, 455--478 (1981)

\end{thebibliography}
\end{document}